# Precision Measurement of the Newtonian Gravitational Constant Using Cold Atoms

G. Rosi[1], F. Sorrentino[1], L. Cacciapuoti[2], M. Prevedelli[3] & G. M. Tino[1]

[1]*Dipartimento di Fisica e Astronomia and LENS, Università di Firenze - INFN Sezione di Firenze, Via Sansone 1, 50019 Sesto Fiorentino, Italy*

[2]*European Space Agency, Keplerlaan 1 - P.O. Box 299, 2200 AG Noordwijk ZH, The Netherlands*

[3]*Dipartimento di Fisica, Università di Bologna, Via Irnerio 46, 40126, Bologna, Italy*

**About 300 experiments tried to determine the value of the Newtonian gravitational constant $G$ to date but large discrepancies in the results prevent from knowing its value precisely[1]. The weakness of the gravitational interaction and the impossibility of shielding the effects of gravity make it very difficult to measure $G$ keeping systematic effects under control. Most of the experiments performed so far were based on the torsion pendulum or torsion balance scheme as in the experiment by Cavendish[2] in 1798 and in all cases they used macroscopic masses. Here we report on the first precise determination of $G$ using laser-cooled atoms and quantum interferometry to probe gravity. We obtain the value $G= 6.67191(99) \times 10^{-11}\,\mathrm{m^3\,kg^{-1}\,s^{-2}}$ with a relative uncertainty of 150 ppm. Our value is at 1.5 combined standard deviations from the current recommended value of the Committee on Data for Science and Technology (CODATA)[3]. Such a conceptually different experiment is important to identify the so far elusive systematic errors thus improving the confidence in the value of $G$. There is no definitive relationship indeed between $G$ and the other fundamental constants and no theoretical pre-**

**diction for its value to test the experimental results. Improving the knowledge of $G$ has not only a pure metrological interest, but is also important for the key role that this fundamental constant plays in theories of gravitation, cosmology, particle physics, astrophysics, and geophysical models.**

The basic idea of our experiment is to use an atom interferometer as gravity sensor and a well-characterized mass as source of a gravitational field. From the precise measurement of the atoms' acceleration produced by the source mass and from the knowledge of the mass distribution, it is possible to extract the value of the gravitational constant $G$ using the well-known formula

$$\mathbf{F}(\mathbf{r}) = -G\,\frac{M_1 M_2}{r^2}\hat{\mathbf{r}} \qquad (1)$$

Atom interferometers[4,5] are new tools for experimental gravitation as, for example, in precision measurements of gravity acceleration[6] and gravity gradients[7], as gyroscopes based on the Sagnac effect[8], for testing the $1/r^2$ law[9], general relativity[10] and quantum gravity models[11], and for applications in geophysics[12]. Proof-of-principle experiments to measure $G$ using atom interferometry were reported[13–15]. Ongoing studies show that future experiments in space will allow to take full advantage of the potential sensitivity of atom interferometers for fundamental physics tests[16]. The possibility of using atom interferometry for gravitational waves detection is being studied[17].

Since the problem in the determination of $G$ depends on the presence of unidentified systematic errors, our experiment was designed with a double-differential configuration in order to be as immune as possible from such effects: the atomic sensor was a double interferometer in a gravity-gradiometer configuration, in order to subtract common-mode spurious signals, and we used two sets of well-characterized tungsten masses as source of the gravitational field that were placed in two different positions in order to modulate the relevant gravitational signal. An additional cancellation of common-mode spurious effects was obtained by reversing the direction of the two-photon recoil used to split and recombine the wave packets in the interferometer[18]. Efforts were devoted to the control of systematics related to atomic trajectories, positioning of the atoms, and effects due to stray fields. The high density of tungsten was instrumental to maximize the signal and to compensate the Earth's gravity gradient in the atom interferometers regions thus reducing the sensitivity to the vertical position and size of the atomic probes.

The atom interferometer is realized using light pulses to stimulate $^{87}$Rb atoms on the two-photon Raman transition between the hyperfine levels $F$=1 and $F$=2 of the ground state[19]. The light field is generated by two counter-propagating laser beams with wave vectors $\mathbf{k}_1$ and $\mathbf{k}_2 \simeq -\mathbf{k}_1$ aligned along the vertical direction. The gravity gradiometer consists of two vertically separated atom interferometers operated in differential mode. Two atomic clouds launched along the vertical direction are simultaneously interrogated by the same $\pi/2$-$\pi$-$\pi/2$ pulse sequence. The difference of the phase shifts detected at the output of each interferometer provides a direct measurement of the differential acceleration induced by gravity on the two atomic samples. In this way, any spurious acceleration induced by vibrations or seismic noise on the common reference frame identified by the vertical Raman beams is efficiently rejected.

Fig. 1 shows a schematic of the experiment. The atom interferometer apparatus and the source masses assembly are described in detail elsewhere[20,21]. In the vacuum chamber at the bottom of the apparatus, a magneto-optical trap (MOT) collects $\sim 10^9$ rubidium atoms. After turning the MOT magnetic field off, the atoms are launched vertically along the symmetry axis of the vacuum tube by using the moving molasses technique. During the launch sequence, atoms are laser cooled to a temperature of $\sim 4\,\mu$K. We juggle two atomic samples to have them reach the apogees of their ballistic trajectories at about 60 cm and 90 cm above the MOT with a vertical separation of 328 mm.

The atoms are velocity selected and prepared in the magnetic-field-insensitive $F$=1, m$_F$=0 hyperfine state with a combination of three $\pi$ Raman pulses and two resonant laser pulses used to blow away the atoms occupying the wrong hyperfine state. The interferometers are realized at the center of the vertical tube shown in Fig. 1. In this region, surrounded by two cylindrical magnetic shields, a uniform magnetic field of $29\,\mu$T along the vertical direction defines the quantization axis. Here, atoms are subjected to the Raman three-pulse interferometer sequence. The central $\pi$ pulse occurs about 6 ms before the atoms reach the apogees of their trajectories. At the end of their ballistic flight, the population of the ground state is measured by selectively exciting the atoms in both hyperfine levels of the ground state and detecting the light-induced fluorescence emission. We typically detect $10^5$ atoms on each rubidium sample at the end of the interferometer sequence. Each measurement takes 1.9 s. The information on the relative phase shift between the two atom interferometers is obtained from a Lissajous plot of the signal of one interferometer vs the other. Experimental points distribute along an ellipse. The differential phase shift is extracted from the eccentricity and the rotation angle of the ellipse fitting the data[22]. The instrument sensitivity for differential acceleration is $3 \times 10^{-9}$ g for 1 s of integration.

The source mass is composed of 24 tungsten alloy cylinders, for a total mass of about 516 kg. Each cylinder is machined to a diameter of 99.90 mm and a height of 150.11 mm. They are positioned on two titanium platforms and distributed with hexagonal symmetry around the vertical axis of the tube (see Fig. 1). The cylinders' centers lie along two circles with nominal radii $2R$ and $2R\sqrt{3}$ respectively, where $R$ is the radius of a single cylinder. The vertical positioning of the two platforms is ensured by precision screws synchronously driven by stepper motors and by an optical readout system. The reproducibility of the positioning was verified with a laser tracker to be within 1 $\mu$m[20]. With respect to the position of the apogee of the lower atomic cloud, the centers of the lower and upper set of cylinders lie at a vertical distance of 40 mm and 261 mm respectively for the C configuration and -74 mm and 377 mm respectively for the F configuration.

The value of the Newtonian gravitational constant was obtained from a series of gravity gradient measurements performed by periodically changing the vertical position of the source masses between configuration $C$ and $F$. Fig. 2 shows the data used for the determination of $G$. Data were collected in 100 hours during one week in July 2013. Each phase measurement was obtained by fitting a 360-point scan of the atom interference fringes to an ellipse. The modulation of the differential phase shift produced by the source mass is well visible and can be resolved with a signal-to-noise ratio of 1000 after about one hour. The resulting value of the differential phase shift is 0.547870(63) rad from which the $G$ value was obtained. The measured differential phase shift is produced for 97.0% by the cylinders, for 2.8% by the cylinders supports[20], and for 0.2% by the additional moving masses (translation stages, optical rulers, screws).

A budget of the uncertainty sources affecting the value of $G$ is presented in Table 1. Positioning errors account for uncertainties in the positions of the 24 tungsten cylinders along the radial and vertical direction, both in configuration $C$ and $F$. Density inhomogeneities of the source masses were measured by cutting and weighing a spare cylinder[20] and modelled in the data analysis. The precise knowledge of the atomic trajectories is of key importance to analyze the experimental results and derive the value of $G$. The velocity of the atomic clouds and their position at the time of the first interferometer pulse were calibrated by time-of-flight measurements and by detecting the atoms when they crossed a horizontal light sheet while moving upward and downward. The Earth rotation affects the atom interferometer signal because of the transverse velocity distribution of the atoms in the fountain. Following the method demonstrated for a single interferometer[23,24], we implemented a tip-tilt scheme for the mirror retroreflecting the Raman beams in our double interferometer.

The analysis to extract the value of $G$ from the data involved the following steps: (1) calculation of the gravitational potential produced by the source masses, (2) calculation of the phase shift for single atom trajectories, (3) Monte Carlo simulation of the atomic cloud, (4) calculation of the corrections for the effects not included in the Monte Carlo simulation which are listed in Table 1.

After an analysis of the error sources affecting our measurement, we obtain the value $G = 6.67191(77)(65) \times 10^{-11}\,\mathrm{m^3\,kg^{-1}\,s^{-2}}$. The statistical and systematic errors, reported in parenthesis as one standard deviation, lead to a combined relative uncertainty of 150 ppm. In Fig. 3, this result is compared with the values of recent experiments and CODATA adjustments. Our value, obtained with a method completely different from the ones used before, is at 1.5 combined standard deviations from the current CODATA value $G = 6.67384(80) \times 10^{-11}\,\mathrm{m^3\,kg^{-1}\,s^{-2}}$.

Such conceptually different experiments will be important to identify errors that produced the still unsolved discrepancies amongst previous measurements. Given the relevance of the gravitational constant in several fields ranging from cosmology to particle physics and in the absence of a complete theory linking gravity to other forces, high precision measurements based on different methods are crucial to improve the confidence in the value of this fundamental constant. It is possible to envisage a new generation of experiments to measure $G$ using ultracold atoms confined in optical traps for their precise positioning. The remaining major contribution to the systematic uncertainty in our experiment derives indeed from the positioning of the atoms with respect to the source mass. The choice of a different atom might also reduce other sources of uncertainties. For example, Sr atoms can be rapidly cooled down to Bose Einstein condensation[25], they are virtually immune to magnetic fields and collisional effects in their ground state, and the possibility of precise positioning and gravity measurements using optical lattices was demonstrated[26,27]. Higher sensitivity atom interferometers with larger splitting of the atomic wave-packet would enable the use of a smaller highly homogeneous source mass such as gold or eventually silicon crystals. We foresee prospects for pushing the measurement accuracy in the determination of the gravitational constant to better than 10 ppm using atoms.

**Methods Summary** The Raman lasers frequencies $\nu_1$ and $\nu_2$ match the resonance condition $\nu_2 - \nu_1 \simeq 6.8$ GHz, corresponding to the separation between the $F$=1 and $F$=2 levels of $^{87}$Rb ground state. During absorption and stimulated emission processes, the resonant light field exchanges with atoms a total momentum of $\hbar\mathbf{k} = \hbar\mathbf{k}_1 - \hbar\mathbf{k}_2$, coupling the two states $|F=1,\mathbf{p}\rangle$ and $|F=2,\mathbf{p}+\hbar\mathbf{k}\rangle$, where $\mathbf{p}$ is the initial atomic momentum. The interferometer is composed of a $\pi/2$-$\pi$-$\pi/2$ sequence of Raman pulses separated by a time $T = 160$ ms. The pulses split, redirect and recombine the wave-packets producing atom interference. At the end of the sequence, the probability of detecting atoms in $|F=2,\mathbf{p}+\hbar\mathbf{k}\rangle$ is $P_2 = \frac{1}{2}[1 - \cos\phi]$, where $\phi$ is the phase difference accumulated along the interferometer arms. In a uniform gravity field, atoms experience a phase shift $\phi = \mathbf{k} \cdot \mathbf{g} T^2$ that can be measured thus providing direct information on the local acceleration due to gravity. Accurate control of noise sources and systematic phase shifts is crucial to optimize the sensitivity and the long-term stability of gravity gradient measurements[28]. Active loops have been implemented to stabilize the optical intensity of cooling, Raman and probe laser beams and the Raman mirror tilt. In this way, we reach a quantum-projection-noise-limited sensitivity to differential accelerations of $3 \times 10^{-9}$ g after 1 s of integration time and an accuracy on the $G$ measurement at the level of 100 ppm. The measured phase shift is compared to the theoretical value obtained by evaluating the action integral along the trajectories given by the solution of the Lagrange equation:

$$\phi = \frac{1}{\hbar} \oint \mathcal{L} dt. \tag{2}$$

Following a perturbative approach[29], we separate the Lagrangian $\mathcal{L}$ into $\mathcal{L}_0 = p^2/(2m) - mgz$, $m$ being the atomic mass, and $\mathcal{L}_1$ accounting for Earth gravity gradient and source masses contribu-tion. A Monte Carlo simulation is used to evaluate the measurement results and to determine the sensitivity of the final phase angle to the relevant parameters. To this purpose, the key parameters entering the simulation are varied and their derivatives calculated to estimate the uncertainty on the $G$ measurement.

**Acknowledgements**  G.M.T. acknowledges seminal discussions with M.A. Kasevich, and J. Faller and useful suggestions by A. Peters in the initial phase of the experiment. We are grateful to A. Cecchetti and B. Dulach of INFN-LNF for the design of the source mass support and to A. Peuto, A. Malengo, and S. Pettorruso of INRIM for density tests on W mass. We thank D. Wiersma for a critical reading of the manuscript. This work was supported by INFN (MAGIA experiment).

**Author Contributions**  G. M. T. conceived the experiment, supervised it and wrote the manuscript. G. R., F. S. and L. C. performed the experiment. M. P. contributed to the experiment and analysed the data.

**Author Information** The authors declare that they have no competing financial interests. Correspondence and requests for materials should be addressed to G. M. T. (guglielmo.tino@fi.infn.it).

**Methods**

**Experimental setup** Our interferometer uses Raman pulses to drive $^{87}$Rb atoms on the two-photon transition between the hyperfine levels $|1\rangle \equiv |F = 1, m_F = 0\rangle$ and $|2\rangle \equiv |F = 2, m_F = 0\rangle$ of the ground state[19]. The light field is generated by two counter-propagating laser beams with wave vectors $\mathbf{k}_1$ and $\mathbf{k}_2 \simeq -\mathbf{k}_1$ aligned along the vertical direction. This configuration is obtained by retro-reflecting the Raman lasers by a mirror placed on top of the vertical tube (see Fig. 1). The laser frequencies $\nu_1$ and $\nu_2$ match the resonance condition $\nu_2 - \nu_1 = \nu_0 + \nu_{AC} + \nu_{rec}$, where $h\nu_0$ is the energy corresponding to the $|1\rangle \to |2\rangle$ transition, $h\nu_{AC}$ accounts for the AC Stark effect and $h\nu_{rec}$ is the energy shift due to the two-photon recoil. As a consequence, the transition modifies both the internal energy and the total momentum of the atom, coupling the two atomic states $|1, \mathbf{p}\rangle$ and $|2, \mathbf{p} + \hbar\mathbf{k}\rangle$, where $\mathbf{k} = \mathbf{k}_1 - \mathbf{k}_2$ is the effective wave vector of the Raman transition. The $\hbar\mathbf{k}$ momentum transfer is responsible for the spatial separation of the atomic wave-packet along the two physical arms of the interferometer. During the interferometer, a sequence of three Raman pulses provides the atom-optical elements of a typical Mach-Zehnder configuration: at $t = 0$ a $\pi/2$ beam splitter pulse prepares the atoms, initially in $|1, \mathbf{p}\rangle$, in an equal and coherent superposition of $|1, \mathbf{p}\rangle$ and $|2, \mathbf{p}+\hbar\mathbf{k}\rangle$; after a time $T = 160$ ms a $\pi$ pulse plays the role of a mirror, swapping state $|1, \mathbf{p}\rangle$ with $|2, \mathbf{p} + \hbar\mathbf{k}\rangle$ and redirecting the atoms towards the output ports of the interferometer; at $t = 2T$ a final $\pi/2$ pulse recombines the atomic wave-packets in the two output ports where they interfere. When interacting with the light field, the phase of the Raman lasers is imprinted on the atomic wave-function, sampling the atomic motion at the three interaction events of the $\pi/2$-$\pi$-$\pi/2$ sequence. This information can be read through the effect of matter-waves interference by detecting the atomic population in the $F = 1$ and $F = 2$ hyperfine states. If $\phi$ is the phase difference accumulated by the atoms along the two interferometer arms, the probability of detecting atoms in $|2\rangle$ can be expressed as $P_2 = \frac{1}{2}[1 - \cos\phi]$. In the presence of a gravity field, atoms experience a phase shift $\phi = \mathbf{k} \cdot \mathbf{g}T^2$. Measuring $\phi$ is therefore equivalent to a measurement of the local acceleration due to the gravity field along the direction of the effective wave vector $\mathbf{k}$.

Our instrument is a gravity gradiometer consisting of two Raman-pulse interferometers simultaneously probing two clouds of laser-cooled $^{87}$Rb atoms aligned along the vertical axis and separated by a distance of 328 mm [21,28]. The two atomic clouds are interrogated by the same Raman lasers with the same interferometric pulse sequence. The difference of the phase shifts at each interferometer provides a direct measurement of the differential acceleration induced by gravity on the two atomic samples. This configuration allows a very efficient rejection of common mode noise sources including spurious accelerations at the instrument platform induced by mechanical vibrations or seismic noise, g variations due to tidal effects and phase noise introduced by the Raman lasers. In addition, the difference of the measurements performed in configurations $C$ and $F$ (see Fig. 1) efficiently rejects long-term systematic drifts which do not depend on the distribution of the source masses. They include wavefront distortions, magnetic field perturbations, Coriolis acceleration and light shift.

Controlling sources of statistical noise and systematic phase shifts on the gravity gradient measurement is crucial for the experiment[28]. Inertial effects induced by Earth's rotation are an important source of both noise and systematic errors. Gravity gradient measurements are indeed sensitive to the Coriolis acceleration when differential velocities along the East-West direction are present between the two atomic samples and when the launch direction itself is not stable. Precise control on the effects of the Coriolis acceleration is achieved by applying a uniform rotation rate to the retro-reflecting Raman mirror by means of PZT actuators during the atom interferometry sequence[23,24]. The optimal rotation rate, counteracting the local projection of the Earth rotation rate on the horizontal plane, is determined to better than 2 $\mu$rad/s after maximizing the interferometers contrast and minimizing the noise on the differential phase measurements. In such conditions, the residual shift amounts to $1.5 \cdot \Delta\theta$, where $\Delta\theta$ is the variation in the launch direction expressed in rad when switching from configuration $C$ to $F$ and viceversa. As a consequence, a $G$ measurement down to 100 ppm level requires a control on the launch direction to better than 37 $\mu$rad. We can control $\Delta\theta$ to better than 8 $\mu$rad. The velocity distribution of the atoms due to the finite temperature of the sample ($\sim 4$ $\mu K$) also contributes a systematic shift on the $G$ measurement. In our experiment, the Raman mirror tilt is controlled to better than 100 nrad bringing the corresponding error on the $G$ measurement to negligible levels. Magnetic fields can perturb atom interferometry measurements both through their mechanical action on the atomic trajectories and via the second-order Zeeman effect. A $10^3$ shielding factor to external magnetic fields is achieved in the interferometer volume by passively isolating the vertical tube through a system of two cylindric $\mu$-metal layers. In addition, systematic shifts that slowly change over a typical time scale of a few seconds and that do not depend on the direction of the effective wave vector **k** are efficiently cancelled by reversing **k** with a frequency shift of the Raman lasers at each measurement cycle and taking the average phase shift[18]. They include the second-order Zeeman effect as well as the first-order light shift. Finally, the main parameters of the experiment have been actively stabilized to optimize the sensitivity and the long-term stability of gravity gradient measurements: the optical intensity of cooling, Raman and probe laser beams are regulated by acting on the RF power driving the acousto-optical modulators; the Raman mirror tilt is controlled by a piezo tip/tilt system. In this way, we obtain a sensitivity to differential accelerations of $3 \times 10^{-9}$ g for an integration time of 1 s, in agreement with the calculated quantum projection noise limit for $2 \times 10^5$ atoms[28].

**Monte Carlo simulation and measurement systematics** Gravity gradient measurements are compared to a numerical simulation that evaluates the gravitational potential produced by a given configuration of the source masses, implements the calculation of the phase shift experienced by single atoms at the two interferometers and finally runs a Monte Carlo simulation on the atomic trajectories by varying initial position and velocity out of the density and velocity distribution of the atomic cloud best fitting the experimental density profiles. The gravitational potential generated by the source masses is computed analytically using a multipole expansion. The formula has been ver-ified by calculating the potential of a single cylindrical mass along the classical path of the atoms during the interferometric sequence and comparing it against the numerical evaluation of the exact solution. The multipole expansion has been found to introduce negligible error while drastically reducing the computation time. The phase shift $\phi$ for an atom is obtained as $1/\hbar$ times the classical action, evaluated along the integration paths given by the solution of the Lagrange equation (Eq. 2). We use a perturbative method[29] separating the Lagrangian $\mathcal{L}$ into $\mathcal{L}_0 = p^2/(2m) - mgz$, $m$ being the atomic mass, and $\mathcal{L}_1$ accounting for the contributions of the Earth's gravity gradient and of the source masses. The phase shift due to $\mathcal{L}_1$ is then calculated by integration over the unperturbed path given by the solution of the Lagrange equation for $\mathcal{L}_0$. This approximation gives a negligible error in the $G$ determination since the ratio $|\mathcal{L}_1/\mathcal{L}_0|$ computed along the unperturbed paths is lower than $10^{-7}$ in our experiment. The validity of the perturbative approach was additionally checked by comparison with the exact solution for the case of a uniform gradient[30]. The interaction processes between the atom and the Raman laser beams during the interferometry sequence are modelled in the code to account for the finite duration of the Raman pulses. At detection, we evaluate the convolution effects due to both the finite width of the detection beams and the finite bandwidth of the photodiodes collecting the fluorescence light at the $F$=1 and $F$=2 channels. Crucial for the Monte Carlo simulation is the spatial and momentum distribution used to describe the atomic sample and its evolution through the velocity selection pulses and the atom interferometer pulse-sequence. Different techniques have been used to characterize the density and velocity profile of the atomic clouds. In particular, time-of-flight measurements provide the required information along the vertical direction, while the profiles in the horizontal plane are measured both by performing CCD imaging and by scanning the atomic cloud using the vertical Raman laser beams with a reduced diameter of 4 mm. The experimental data show that the spatial and velocity profiles of atomic clouds at launch, right after release from the magneto-optical trap, are well described by Gaussian distributions with spatial deviation $\sim 3$ mm, velocity deviation $\sim 16$ mm/s for the lower cloud, and $\sim 22$ mm/s for the upper cloud. The velocity selection during the interferometer reduces the vertical velocity deviation to $\sim 3$ mm/s and the horizontal velocity deviation to $\sim 6$ mm/s. Average transverse velocities are below 1.5 mm/s.

The Monte Carlo simulation is used not only to evaluate the results of our measurements, but also to determine the sensitivity of the final phase angle to the relevant parameters and finally estimate the uncertainty on the $G$ measurement. Derivatives are computed with respect to 28 parameters including the positions of the source masses, positions and velocities of the upper and lower clouds, their time-of-flight at detection, the Gaussian widths of the detection lasers and their position with respect to the atomic clouds, the interferometry sequence duration $2T$, the effective wave vector $k$ of the Raman lasers, the gravitational acceleration and its gradient. From an analysis of the results, the spread of the atomic sample along the vertical direction appears to be responsible for the dominant contribution, requiring a measurement of the Gaussian width for both the upper and lower clouds to better than 0.1 mm. The determination of the atomic trajectories requires a good knowledge of the local gravitational acceleration and of the gravity gradient. Gravity acceleration was measured by using an absolute gravimeter to an uncertainty that contributes with a negligible error to the $G$ measurement. A recent measurement of the gravity gradient performed with our atom interferometer gave $3.11 \times 10^{-6}\,\mathrm{s}^{-2}$, with an effect of less than 2 ppm on the determination of the Newtonian gravitational constant. The $G$ measurement has been corrected for the effect of the finite density of the volume of air that replaces the source masses when they are moved from the configuration $C$ to $F$ and viceversa. The corresponding correction on the $G$ value amounts to $(+60\pm6)$ ppm. Finally, a correction has been applied to account for the non-uniform density of the tungsten cylinders[20]; the axial density inhomogeneity of the cylinders gives the main contribution corresponding to a correction on the $G$ value of $(+90 \pm 20)$ ppm.

The uncertainty budget of our measurement of the Newtonian gravitational constant is detailed in Table 1. Our result is reported in Fig. 3 and compared with the CODATA value[3] and recent determinations for G[31–43].

**Figure 1** **Schematic of the experiment showing the Rb atom interferometer operated as a gravity gradiometer and the W masses used as source of the gravitational field.** For the measurement of $G$, the position of the source masses is alternated between configurations $F$ and $C$. Plots of gravity acceleration produced along the symmetry axis by the source masses are also shown; a constant value for Earth's gravity was subtracted. The spatial regions of the upper and lower atom interferometers are indicated by the thick lines. The vertical acceleration plots show the effect of source masses to cancel the local gravity gradient at the position of the atomic apogees.

**Figure 2** **Experimental data.** a) Typical Lissajous figures obtained by plotting the output signal of the upper atom interferometer vs the lower one for the two configurations of the source masses. b) Modulation of the differential phase shift for the two configurations of source masses for a given direction of the Raman beams $k$ vector. Each phase measurement is obtained by fitting a 360-point scan of the atom interference fringes to an ellipse. The error bars, not visible at this scale, are given by the standard error of the least-squares fit to the ellipse. c) Results of the measurements to determine $G$. Each point is obtained by averaging the signals recorded for each of the two directions of the Raman $k$ vector (see Methods). The data acquisition for each point took about one hour. These data were recorded in different days spanning over one week in July 2013. The error bars are given by the combined error on the angles for the four ellipses. d) Histogram of the data in c).

**Figure 3** **Result of this experiment for the Newtonian gravitational constant $G$ compared with the values obtained in previous experiments and with the recent CODATA adjustments.** Only the experiments considered for the current CODATA 2010 value are included and the successive BIPM-13 result. For details on the experiments and their identification with the acronyms used in the figure, see ref. [3] and the additional references in the Methods section.

Table 1: **Effects, relative corrections and uncertainties considered for the determination of $G$ in this experiment**. Uncertainties are quoted as one standard deviation. The third column contains the corrections we applied due to the effects not included in the Monte Carlo simulation. The bias and systematic error from ellipse fitting are evaluated by a numerical simulation on synthetic data. Other effects include: cylinders mass, cylinders vertical position, gravity gradient, gravity acceleration, Raman mirror tilt, Raman $k$ vector, timing.

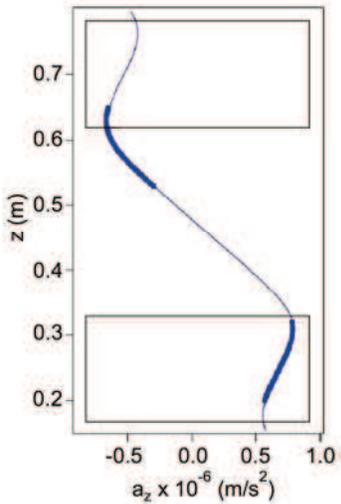 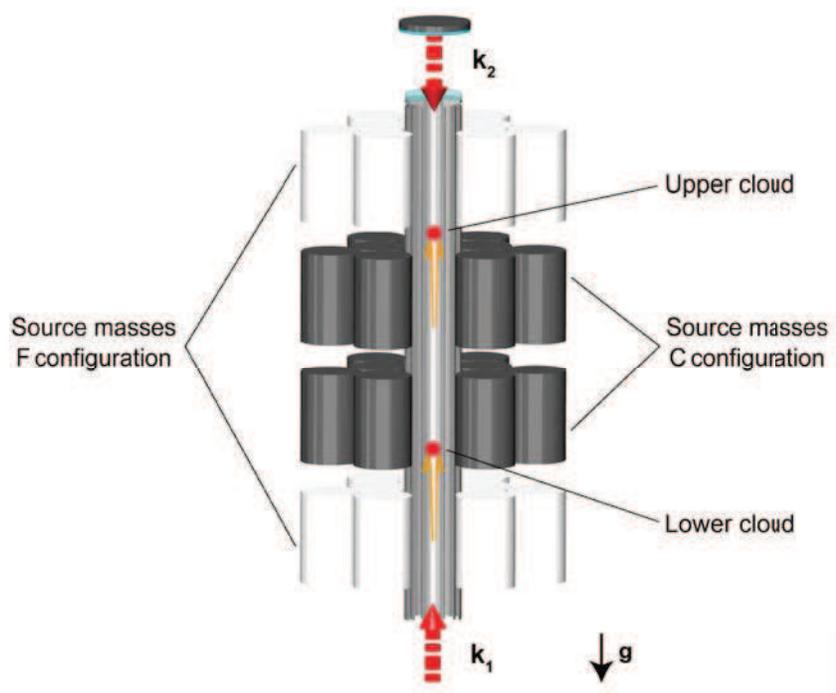 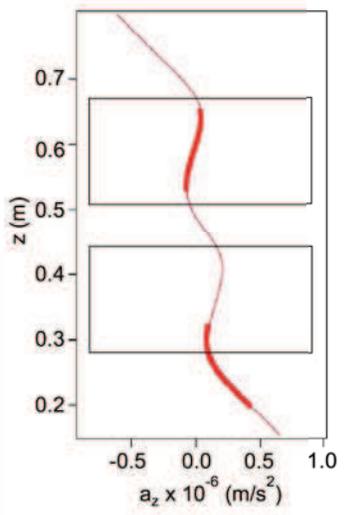

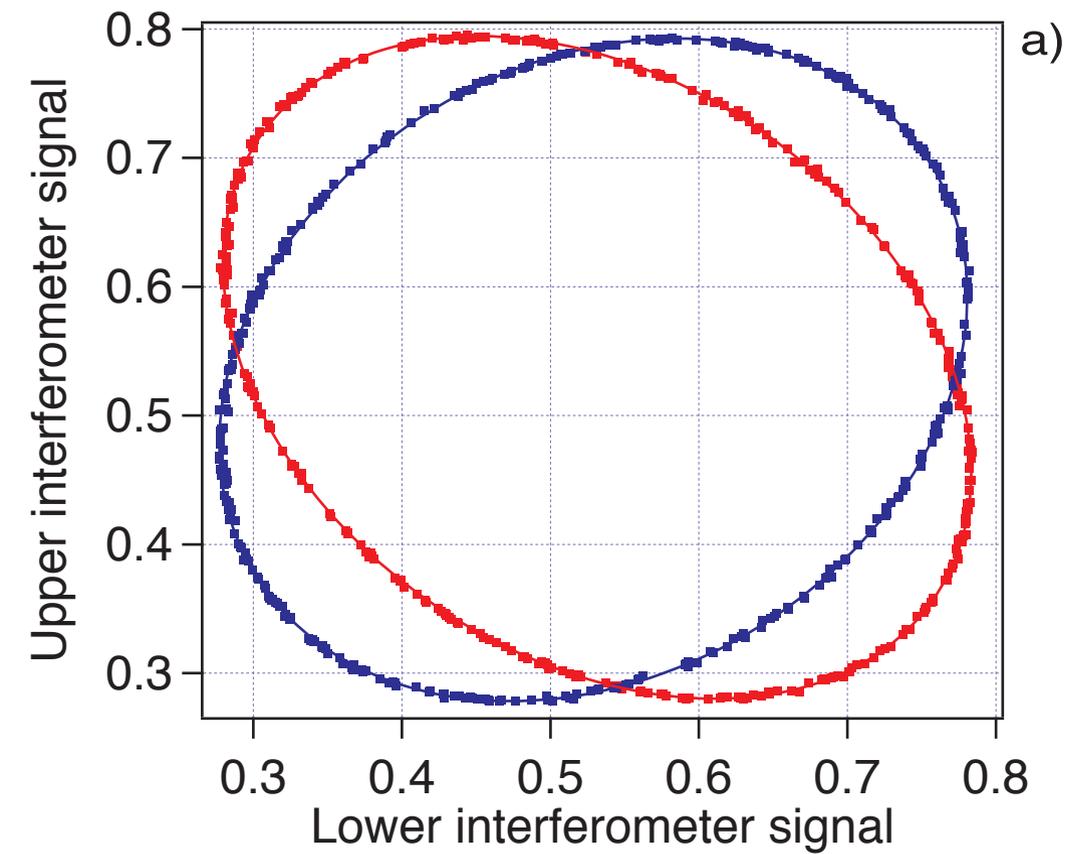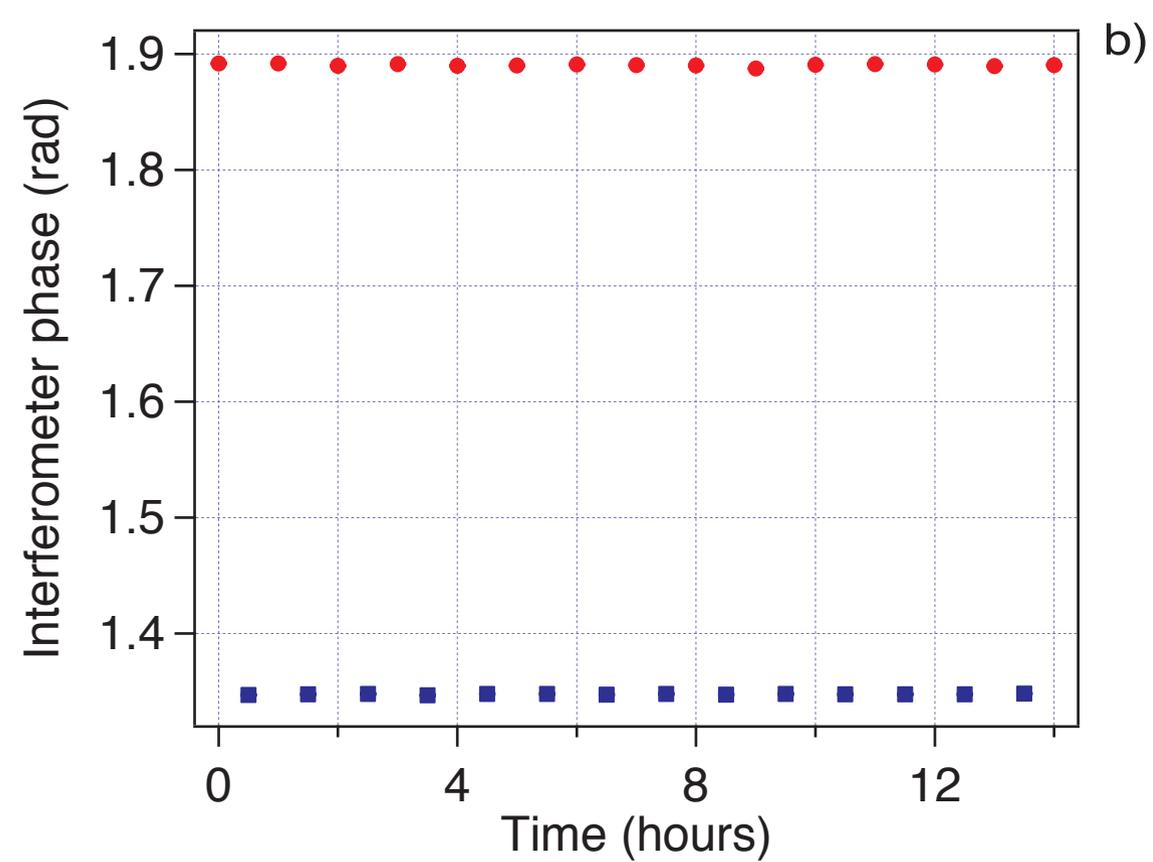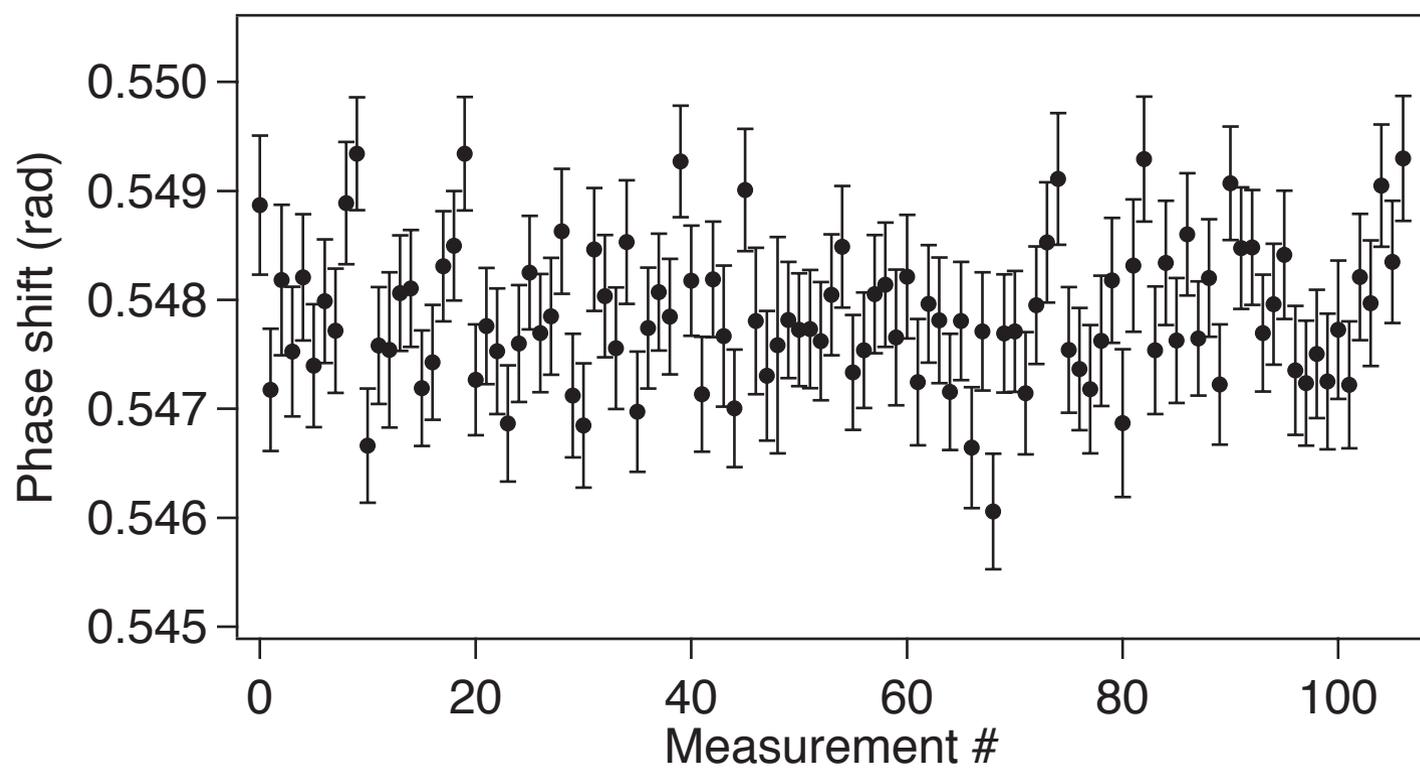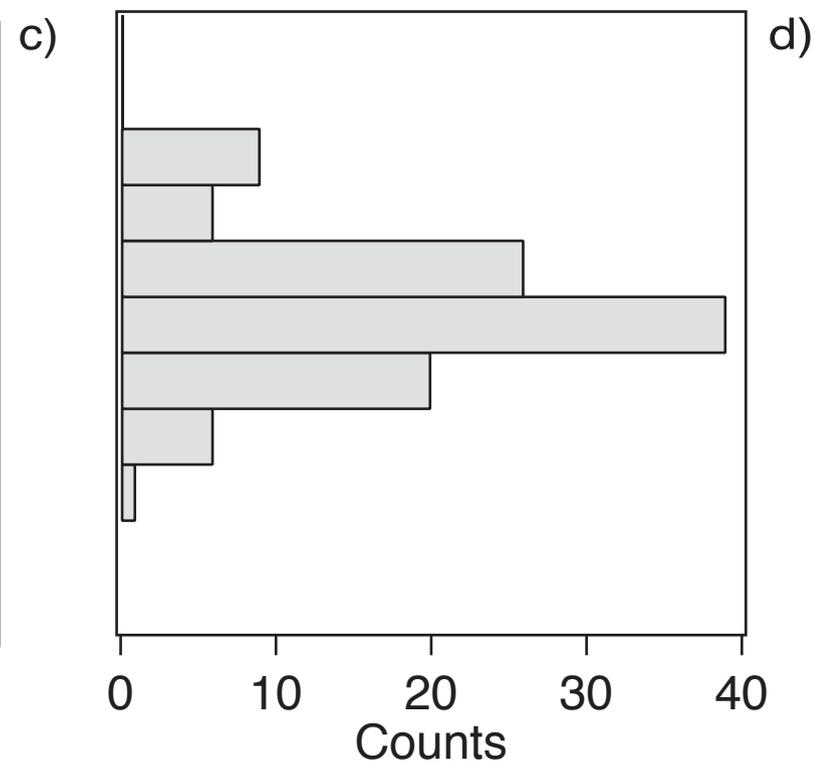

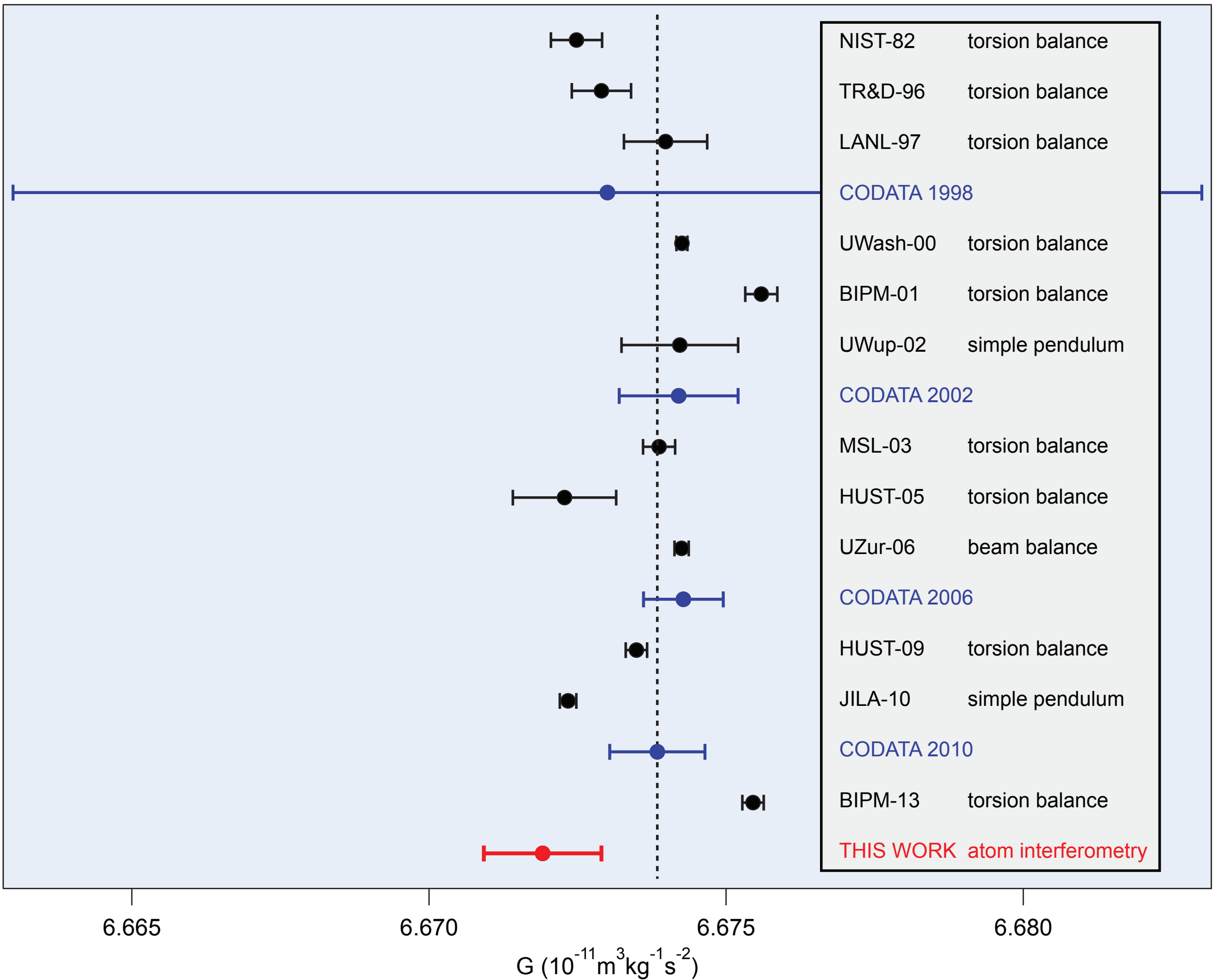

| Effect | Uncertainty | Correction to $G$ (ppm) | Relative uncertainty $\Delta G/G$ (ppm) |
|---|---|---|---|
| Air density | 10 % | 60 | 6 |
| Apogee time | 30 $\mu$s | | 6 |
| Atomic clouds horizontal size | 0.5 mm | | 24 |
| Atomic clouds vertical size | 0.1 mm | | 56 |
| Atomic clouds horizontal position | 1 mm | | 37 |
| Atomic clouds vertical position | 0.1 mm | | 5 |
| Atoms launch direction change C/F | 8 $\mu$rad | | 36 |
| Cylinders density inhomogeneity | $10^{-4}$ | 91 | 18 |
| Cylinders radial position | 10 $\mu$m | | 38 |
| Ellipse fitting | | -13 | 4 |
| Size of detection region | 1 mm | | 13 |
| Support platforms mass | 10 g | | 5 |
| Translation stages position | 0.5 mm | | 6 |
| Other effects | | <2 | 1 |
| Systematic uncertainty | | | 92 |
| Statistical uncertainty | | | 116 |
| Total | | 137 | 148 |